\documentstyle[amssymb,preprint,aps]{revtex}

\begin{document}
\title{{\bf Horizon Dynamics of Evaporating Black Holes in a Higher Dimensional
Inflationary Universe}}
\author{Manasse R. Mbonye}
\address{{\it Physics Department, University of Michigan, Ann Arbor, Michigan 48109}}
\maketitle

\begin{abstract}
Spherically symmetric Black Holes of the Vaidya type are examined in an
asyptotically de Sitter,\ higher dimensional spacetime. The various horizons
are identified and located. The structure and dynamics of such horizons are
studied.
\end{abstract}

\begin{center}
PACS number(s):\ 04.70.Nr, 04.50.+h, 04.70.Bw, 04.40.Dy

\ 
\end{center}

\section{Introduction\protect\bigskip}

The notion that our universe may have begun from a hot big-bang forms the
pillar of the standard cosmological model, a model that plots the time
evolution of the universe. This model holds that soon after the big-bang $%
\left( \thicksim 10^{-36}s\right) $ the temperature of the universe drops
below the critical grand-unified theory temperature $\left( T_{GUT}\thicksim
10^{14}GeV\right) $. In this over-cool sub-stable state, the universe is
dominated by vacuum energy-density with an effective cosmological constant 
\[
\Lambda =\frac{2}{15}\pi ^{3}\left( \frac{T_{GUT}^{2}}{m_{p}}\right) ^{2}
\]
where $m_{p}$ is the Planck mass $\left( \thicksim 10^{19}\,GeV\right) $.
This energy then drives the same universe into a de Sitter-like exponential
inflation $\left[ 1\right] $.\ The standard cosmological model has been very
successful in explaining the origins of most of the properties of our
universe: from the very small to the very big, from nucleosynthesis to large
scale structure formation and the expansion of the universe.

This standard model of cosmology has not, however, been able to address the
physics of the events prior to the inflationary era. Understanding this very
short but very important era of the universe requires a theory that
incorporates General Relativity and Quantum Theory. Such an effort
translates into unifying gravity with the other three forces of nature.
Judging by what has been learned from such candidates as Supergravity $\left[
2\right] $ and now Superstrings $\left[ 3\right] $, it appears that such a
unifying theory will likely require the universe to be multi-dimensional. In
Superstrings, for example, the view is that the universe started out in a
higher-dimensional phase. Four of the dimensions then expand leaving behind
the rest in the regime of the planck length. In the field theoretic-limit of
Superstring theories, gravity is described reasonably accurately by
multi-dimensional Einstein field equations $\left[ 3\right] $. In this
unification scheme, internal symmetries can be traced to the spacetime
associated with the extra dimensions $\left[ 4\right] $. Gauge invariance
then assumes the same status as spacetime invariance and internal quantum
numbers such as electric charge, believed to result from symmetry motions in
the extra dimensions, are now brought to the same footing as energy and
momentum.

In searching for possible effects of the extra dimensions it makes sense,
therefore, to evolve the Friedman-Robertson-Walker model of our universe
back in time to the early de Sitter phase where one reaches energies at
which such extra dimensions may be resolvable (see [4] and Refs. therein).
Coincidentally, this too turns out to be the era when primordial black holes
may have been produced $\left[ 5-9\right] $. For this \ and other reasons
(like the distinctly high curvature nature of the spacetime around them)
black holes will continue to be important probes in any quantum theory of
gravity, and in the quest to understand the role of multi-dimensionality in
the early (and possibly present? (see $\left[ 10\right] $)) evolution of our
universe. The recent findings of links between black hole entropy and
topological structures of the extra dimensions (see for example $\left[ 11%
\right] $ and $\left[ 12\right] $) not only puts high-dimensionality to a
stronger footing but also emphasizes the role black holes will play as
probes in such future research. Naturally this calls for an understanding of
the dynamics of black holes in higher dimensions, particularly under
conditions that mimic those of \ the early universe. In this paper we begin
a study of\ the dynamics of such a black hole in such a setting.

Several solutions to the Einstein equations of localized sources in higher
dimensions have been obtained in the recent years. This includes the higher
dimensional generalizations of the Schwarzschild and the Reisner-Nordstrom
solutions $\left[ 13,14\right] $, the Kerr solution $\left[ 15,16\right] $
and the Vaidya solution $\left[ 17\right] $. Recently the metric of a
radiating black hole in a de Sitter background, that is a generalization of
the Mallett $\left[ 18\right] $ metric, has been written down $\left[ 19,20%
\right] $. In the present work our aim is to demonstrate that the dynamics
of a radiating black hole in a higher dimensional cosmological background
can be sensibly discussed. First, we seek to identify and locate the various
horizons. After this we then go on to study the structures and discuss the
dynamics of such horizons. It is shown, at each stage, that all the results
we obtain reduce to the well known Mallett $\left[ 21\right] $ results as we
go down to four dimensions, and to make this transparent our analysis is
closely modelled to that of Mallett.

In Section II we introduce the working metric and the theoretical
background. In Section III we derive equations for the horizons. We solve
these equations and use the solutions to identify and locate the various
horizons in the problem. In Section IV we take up the issue of the structure
of such horizons and study their dynamics. In Section V we conclude the
discussion.

\section{The metric and the theory}

\subsection{The Metric}

In this treatment we wish to consider a radiating black hole introduced in
an N dimensional de Sitter space-time. We suppose, for simplicity, that such
a black hole is reasonably modelled by an imploding shell of
negative-energy-density null fluid in the de Sitter universe. In advanced
time, comoving, coordinates the metric [19, 20] is given by the line element
,

\begin{equation}
ds^{2}=-\left[ 1-\frac{2G_{N}m\left( v\right) }{nr^{n}}-\frac{2\Lambda }{%
\left( n+1\right) \left( n+2\right) }r^{2}\right] dv^{2}+2dvdr+d\Omega
_{n+1}^{2}  \eqnum{2.1}
\end{equation}
where $n=N-3$, $m\left( v\right) $, the mass, is a monotonically decreasing
function of the advanced time coordinate $v$, $G_{N}$ is the N-dimensional
gravitational constant, $\Lambda $ is the cosmological constant and 
\begin{equation}
d\Omega _{n+1}^{2}=r^{2}\left( d\theta _{1}^{2}+\sin ^{2}\theta _{1}d\theta
_{2}^{2}+.....+\sin ^{2}\theta _{1}\sin ^{2}\theta _{2}....\sin ^{2}\theta
_{n}d\theta _{n+1}^{2}\right)  \eqnum{2.2}
\end{equation}
\ is the line element on the $\left( n+1\right) $-sphere. For N = 4 the
geometry reduces to that of the Vaidya-Mallett space-time [18].

In the absence of the cosmological background, $\Lambda $, the usual
luminosity, $L_{0}$, of the black hole is defined [19] from the only
non-vanishing component of the energy-momentum tensor, 
\begin{equation}
T_{v}^{r}=T_{vv}=\frac{\left( n+1\right) }{8\pi n}\frac{\dot{m}\left(
v\right) }{r^{n+1}}  \eqnum{2.3}
\end{equation}
as 
\begin{equation}
L_{0}=-\dot{m}\   \eqnum{2.4}
\end{equation}
Here and henceforth $\dot{m}\left( v\right) $ and $m^{\prime }\left(
v\right) $ denote, as usual, derivatives with respect to the time and space
coordinates, respectively. The Luminosity which is bounded from above, $%
L_{0}<1$, is measured in regions where $\frac{d}{dv}$ is time-like.

One can introduce a basis of vectors at every point in this spacetime. \ Two
such vectors $\beta _{a}$ and $l_{a}$ span the radial-temporal subspace and
are given by 
\begin{equation}
\beta _{a}=\delta _{a}^{v}  \eqnum{2.5}
\end{equation}
and 
\begin{equation}
l_{a}=-\frac{1}{2}\left[ 1-\frac{2G_{N}m\left( v\right) }{nr^{n}}-\frac{%
2\Lambda }{\left( n+1\right) \left( n+2\right) }r^{2}\right] \delta
_{a}^{v}+\delta _{a}^{r}  \eqnum{2.6}
\end{equation}
while the rest of the $\left( N-2\right) $ vectors are defined on the $n+1$%
-sphere and induce on the latter a tensor field $\gamma _{ab}$ of the form 
\begin{equation}
\gamma _{ab}=r^{2}\left( \delta _{a}^{\theta _{1}}\delta _{b}^{\theta
_{1}}+\sin ^{2}\theta _{1}\delta _{a}^{\theta _{2}}\delta _{b}^{\theta
_{2}}+.....+\sin ^{2}\theta _{1}\sin ^{2}\theta _{2}....\sin ^{2}\theta
_{n}\delta _{a}^{\theta _{n+1}}\delta _{b}^{\theta _{n+1}}\right) . 
\eqnum{2.7}
\end{equation}
\ \ \ The vectors satisfy the conditions

\begin{equation}
\beta _{a}\beta ^{a}=l_{a}l^{a}=0,\;\gamma _{ab}\beta ^{b}=\gamma
_{ab}l^{b}=0,\;\beta _{a}l^{a}=-1.  \eqnum{2.8}
\end{equation}

One can do a null-vector decomposition of the above metric in this basis so
that

\begin{equation}
g_{ab}=-\beta _{a}l_{b}-l_{a}\beta _{b}+\gamma _{ab}.  \eqnum{2.9}
\end{equation}

\subsection{Deformation of relativistic Membranes}

The structure and dynamics of horizons of such non-static metrics can be
approached from the non-perturbative description of deformation of
relativistic membranes. In general, one considers the evolution of such
deformations of an arbitrary D-dimensional membrane in an arbitrary
N-dimension space-time. A significant amount of literature has been written
on the subject of such deformations [22, 23, 24]. The quantities that
characterize how a variation in the symmetry of a membrane evolves are the
expansion rate, $\theta $, the shear rate, $\sigma $, and the vorticity
(twist), $\omega $. For the general D-dimensional membrane case, the
equations for the deformations have usually yielded no simple clear
interpretation of these quantities [23]. Recently, Zafiris [24] has made
some progress in addressing the problem. However, for the special $D=1$ case
in an $N$ dimensional background one can still generalize the Carter [22]
form of the Rachaudhuri [25] equation so that

\begin{equation}
\frac{d\theta }{dv}=\kappa \theta -\left( \gamma _{c}^{c}\right) ^{-1}\theta
^{2}-\sigma _{ab}\sigma ^{ab}+\omega _{ab}\omega ^{ab}-R_{ab}l^{a}l^{b} 
\eqnum{2.10}
\end{equation}
with $\theta $, $\sigma $, and $\omega $ taking their usual physical
meaning. Here $R_{ab}$ is the N-dimensional Ricci tensor, $\gamma _{c}^{c}$
is the trace of the projection tensor for null geodesics and $\kappa $ is to
be identified as the surface gravity. The latter is given by 
\begin{equation}
\kappa =-\beta ^{a}l^{b}\nabla _{b}l_{a},  \eqnum{2.11}
\end{equation}
where $\nabla $ is the covariant derivative operator. For infinitesimally
neighboring members of the congruence separated by a relative separation
vector $d{\bf x}$ the rate of the change of separation is given [22]\ by 
\begin{equation}
\frac{1}{2}(ds^{2}\dot{)}=\theta _{ab}dx^{a}dx^{b}.  \eqnum{2.12}
\end{equation}
The expansion rate $\theta $, of a null geodesic congruence is then given by
the trace of the expansion tensor as 
\begin{equation}
\theta =\theta _{a}^{a}=\gamma ^{ab}\nabla _{a}l_{b}.  \eqnum{2.13}
\end{equation}
It follows then that 
\begin{equation}
\nabla _{a}l^{a}=\frac{\partial l_{a}}{\partial x^{a}}+\Gamma
_{ac}^{a}l^{c}=\kappa +\theta .  \eqnum{2.14}
\end{equation}
And clearly in flat space-time $\theta $ vanishes since the connection
coefficients $\Gamma _{ac}^{a}$ will.

\section{Location of the Horizons}

Spherically symmetric irrotational space-times, such as under consideration,
are vorticity \ and the shear free. The structure and dynamics of the
horizons are then only dependent on the expansion, $\theta $. Following York
[26]\ we note that to $O(L_{0})$ the evolution of an apparent horizons ($AH$%
) is to satisfy the requirement that $\theta \simeq 0$, while that of an
event horizons ($EH$) is to satisfy the requirement that $\frac{d\theta }{dv}%
\simeq 0$.

\subsection{The Apparent Horizons}

We have written the general expression for the expansion $\theta $\ as 
\begin{equation}
\theta =\gamma ^{ab}\nabla _{a}l_{b}  \eqnum{3.1}
\end{equation}
Using equations $\left( 2.1\right) $, $\left( 2.6\right) $ and $\left(
2.7\right) $ in $\left( 3.1\right) $ yields 
\begin{equation}
\theta \left( r\right) =\frac{n+1}{2r}\left[ 1-\frac{2G_{N}m\left( v\right) 
}{nr^{n}}-\frac{2\Lambda }{\left( n+1\right) \left( n+2\right) }r^{2}\right]
.  \eqnum{3.2}
\end{equation}

Consider now the function $f\left( r\right) =-\frac{2}{n+1}r\theta $. Since
the York conditions require that at the ($AHs$) $\theta $ and hence $f$
vanish, it follows from equation (3.2) that these surfaces will satisfy 
\begin{equation}
r^{n+2}-\frac{\left( n+1\right) \left( n+2\right) }{2\Lambda }r^{n}+\frac{%
\left( n+1\right) \left( n+2\right) }{n}\frac{G_{N}m\left( v\right) }{%
\Lambda }=0.  \eqnum{3.3}
\end{equation}
As it stands equation (3.3) will obviously not admit simple closed form
solutions. It is possible, however, to put this equation in a useful form
that yields solutions which for practical purposes can, reasonably and
justifiably, be taken as the working solutions to the problem at hand. To
this end it is helpful to first gain some insight in the nature of the
function $f\left( r\right) $ whose roots satisfy (3.3). One notices that the
turning points for the function $f\left( r\right) $ are located at points
where

\begin{equation}
\left[ r^{2}-\frac{n\left( n+1\right) }{2\Lambda }\right] r^{n-1}=0. 
\eqnum{3.4}
\end{equation}
The derivative $\frac{df}{dr}$ then vanishes at points $\ r=0$ and $r=\pm %
\left[ \frac{n\left( n+1\right) }{2\Lambda }\right] ^{\frac{1}{2}}$. From
this one then notices that between $r=+\infty $ and $r=+\left[ \frac{n\left(
n+1\right) }{2\Lambda }\right] ^{\frac{1}{2}}$ the function changes sign
from positive to negative which implies that the function has a root
imbedded in between these two values. \ One notices further, that between \ $%
r=+\left[ \frac{n\left( n+1\right) }{2\Lambda }\right] ^{\frac{1}{2}}$ and $%
r=0$, $f\left( r\right) $ does again change signs this time from negative to
positive, and this betrays the existence of yet a second root. We then
establish that the function $f\left( r\right) $\ has at least two real
roots. More importantly we establish that for $r>0$ this function has two,
and only two, real roots. This is an encouraging result \ consistent with
our expectations [21] that there should be two apparent horizons $r_{AH}^{-}$
and $r_{AH}^{+}$ associated with the black hole and the cosmological
constant, respectively.

The case of the $r^{n-1}$ turning points at $r=0$ suggests the existence of
other roots for the function $f\left( r\right) $. Such roots are actually
complex and would suggest the existence of {\it ghost horizons} (at least
for an observer in our four dimensional spacetime). I will defer, for now,
further speculation of their significance and other issues about the turning
points for a future discussion. It is, however, interesting to note that
such {\it ghost horizons }only{\it \ }start to appear at five dimensions $%
\left( n=2\right) $ and persist for higher dimensions.

To put equation (3.3) in a more manageable form, it is useful to institute a
change of variables. Thus setting\ 
\begin{equation}
r=k\xi ,  \eqnum{3.5}
\end{equation}
where $k=\sqrt{\frac{\left( n+1\right) \left( n+2\right) }{2\Lambda }}$
equation (3.3) can be cast in a form 
\begin{equation}
\xi ^{n+2}-\xi ^{n}+\beta \left( n\right) =0,  \eqnum{3.6}
\end{equation}
where 
\begin{equation}
\beta \left( n\right) =\frac{2G_{N}m\left( v\right) }{n}\left[ \frac{%
2\Lambda }{\left( n+1\right) \left( n+2\right) }\right] ^{\frac{n}{2}}. 
\eqnum{3.7}
\end{equation}
One notes that since in our model $\Lambda >0$, and so $\beta \left(
n\right) >0$, then (3.6), along with the necessary positive reality of $r$,
imply that $0<\xi <1$ and $0<\beta <1$.

Now set $\xi =1-x$, where $0<x<1$ to cast equation (3.6) in the form 
\begin{equation}
x\left( 2-x\right) \left( 1-x\right) ^{n}-\beta \left( n\right) =0. 
\eqnum{3.8}
\end{equation}

The above limits imposed on $\xi $ and hence $x$ suggest one can seek
approximate solutions through expansion of the quantity $\left( 1-x\right)
^{n}$\ in equation (3.8). Thus keeping in the equation terms up to $x^{4}$
gives 
\begin{equation}
x^{4}+ax^{3}+bx^{2}+cx+d=0,  \eqnum{3.9}
\end{equation}
where 
\[
a=-\frac{6n}{\left( n-1\right) \left( 2n-1\right) },\;b=\frac{6\left(
2n+1\right) }{n\left( n-1\right) \left( 2n-1\right) },\;c=-\frac{12}{n\left(
n-1\right) \left( 2n-1\right) },\; 
\]

and$\;$%
\begin{equation}
d=\frac{6}{n\left( n-1\right) \left( 2n-1\right) }\beta \left( n\right) . 
\eqnum{3.10}
\end{equation}

We should mention that this expansion is strictly valid for $n>2$. The
solutions for the 5-dimensional $\left( n=2\right) $ case are actually
exact. Further, the solutions have the right limits when $n=1$ and for this
case one does recover the exact Mallett's cubic equation [21] for the
4-dimensional case from either of the equations (3.3) or 3.6). In general
this approximation is good in the limit $x\rightarrow 0$, $\xi \rightarrow 1$%
. It, however, breaks down when $x\rightarrow 1$, $\xi \rightarrow 0$. In
this latter limit the solution to (3.6) simply goes to $\xi =\left[ \beta
\left( n\right) \right] ^{\frac{1}{n}}\Longrightarrow r=\left( \frac{2Gm}{n}%
\right) ^{\frac{1}{n}}$and the geometry loses knowledge of the cosmological
constant.

Equation (3.9) has a resolvent cubic equation that can\ be that can be
written in the form 
\begin{equation}
y^{3}+py+q=0  \eqnum{3.11}
\end{equation}
where $p=\frac{1}{3}\left[ \left( 3ac-b^{2}\right) -12d\right] $ and$\;q=%
\frac{1}{27}\left[ 9abc-2b^{3}-27c^{2}-9\left( 3a^{2}+b\right) d\right] $.
Equation (3.11) admits three solutions. One such solution that is real for
the parameters $p$, $q$ as defined above can be written as

\begin{equation}
y_{1}=2\sqrt{\frac{-p}{3}}\cos \frac{1}{3}\varphi ,  \eqnum{3.12}
\end{equation}
where $\frac{\pi }{2}<\varphi <\pi $\ is given by $\varphi =\arccos \left( 
\frac{3q}{2\sqrt{\frac{-p^{3}}{3}}}\right) $.

There are four solutions to the quartic equation (3.9) in $x$. The only
physically interesting solutions to equation (3.9) consistent with $r=k\xi
=k\left( 1-x\right) $ should be real and satisfy $x<1$. There are two such
solutions obtained by letting 
\[
R=\sqrt{\frac{a^{2}}{4}-b+\left( 2\sqrt{\frac{-p}{3}}\right) \cos \frac{1}{3}%
\varphi } 
\]
and 
\begin{equation}
D=\sqrt{\frac{3a^{2}}{4}-R^{2}\left( p,\varphi \right) -2b+\frac{4ab-8c-a^{3}%
}{4R\left( p,\varphi \right) }}  \eqnum{3.13}
\end{equation}
These are: 
\begin{equation}
x_{\pm }=-\frac{1}{4}a+\frac{1}{2}R\left( p,\varphi \right) \pm \frac{1}{2}%
D\left( p,\varphi \right) .  \eqnum{3.14}
\end{equation}
Note that all the time dependency of $R$ and so of $D$ is expressed in $p$
and $\varphi $ via $d\left( \beta \left( m\left( v\right) \right) \right) $.

Using the solutions in equation (3.14) and recalling that $\xi =1-x$ \ we
obtain two values $\xi _{1}$ and $\xi _{2}$.\ On applying these results to
equation (3.5) i.e. $r=k\xi =k\left( 1-x\right) $ we find two solutions $%
r_{1}=r_{AH}^{-}\left( v\right) $ and $r_{2}=r_{AH}^{+}\left( v\right) $
such that

\begin{equation}
r_{AH}^{-}\left( v\right) =r_{1}\simeq k\left\{ 1-\frac{1}{4}\left[ 2D\left(
p,\varphi \right) +2R\left( p,\varphi \right) -a\right] \right\} , 
\eqnum{3.15}
\end{equation}
and 
\begin{equation}
r_{AH}^{+}\left( v\right) =r_{2}\simeq k\left\{ 1+\frac{1}{4}\left[ 2D\left(
p,\varphi \right) -2R\left( p,\varphi \right) +a\right] \right\} 
\eqnum{3.16}
\end{equation}
\ 

In the limit $n\rightarrow 1$ one recovers the well known solutions [21].
Thus 
\begin{equation}
\mathrel{\mathop{\lim }\limits_{n=1}}%
r_{AH}^{-}\left( v\right) =-\left( \frac{2}{\sqrt{\Lambda }}\right) \cos
\left( \frac{1}{3}\Psi +\frac{1}{3}\pi \right)  \eqnum{3.17}
\end{equation}
and 
\begin{equation}
\mathrel{\mathop{\lim }\limits_{n=1}}%
r_{AH}^{+}\left( v\right) =\left( \frac{2}{\sqrt{\Lambda }}\right) \cos
\left( \frac{1}{3}\Psi \right)  \eqnum{3.18}
\end{equation}
where now $\left( \frac{\pi }{2}<\Psi \left( v\right) <\pi \right) =\arccos %
\left[ -3m\left( v\right) \sqrt{\Lambda }\right] $. (Note, however, that $%
\varphi $ is not simply related to $\Psi $ by taking the limit $n\rightarrow
1$). Further, in the limit $\Lambda \longrightarrow 0$ and $n\rightarrow 1$
we recover the black hole apparent horizon $r_{AH}^{-}\left( v\right)
\rightarrow 2m\left( v\right) $ and in the limit $m\rightarrow 0$ and $%
n\rightarrow 1$ we find, as one expects that $r_{AH}^{+}\left( v\right)
\rightarrow \sqrt{\frac{\Lambda }{3}}$. Hence our solutions reduce to all
the well known solutions [21] in a four dimensional spacetime. Consequently,
we identify the locus of $r_{AH}^{-}\left( v\right) $\ in equation (3.15) as
the black hole apparent horizon $\left( AH^{-}\right) $ while we identify
the locus of $r_{AH}^{+}\left( v\right) $ in equation\ (3.16) as the de
Sitter apparent horizon $\left( AH^{+}\right) $, for a black hole in an
N-dimensional background with a cosmological constant.

\subsection{The Event Horizons}

The event horizons are null surfaces. \ To $O(L_{0})$\ the evolution of
these surfaces can be determined from the second of the York conditions that 
$\frac{d\theta }{dv}\simeq 0$. We first show that for a black hole imbedded
in an N-dimensional de Sitter background this condition is satisfied. We
shall then solve the resulting equations to locate the event horizons and
study their structure.

The surface gravity $\kappa $ in this spacetime is given from (2.11) by 
\begin{equation}
\kappa =\frac{G_{N}m\left( v\right) }{r^{n+1}}-\frac{2\Lambda }{\left(
n+1\right) \left( n+2\right) }r.  \eqnum{3.17}
\end{equation}
\bigskip Consider now the acceleration of the geodesics $\frac{d^{2}r}{dv^{2}%
}$ \ in our space-time, parametrized by $v$. Since from the line element
(2.1) we have that 
\begin{equation}
\frac{dr}{dv}=\frac{1}{2}\left[ 1-\frac{2G_{N}m\left( v\right) }{nr^{n}}-%
\frac{2\Lambda }{\left( n+1\right) \left( n+2\right) }r^{2}\right] 
\eqnum{3.18}
\end{equation}
then 
\begin{equation}
\frac{d^{2}r}{dv^{2}}=\frac{G_{N}L_{0}}{nr^{n}}+\kappa \frac{dr}{dv} 
\eqnum{3.19}
\end{equation}
Equations (3.2) and (3.18) in (3.19) give 
\begin{equation}
\frac{d^{2}r}{dv^{2}}=\frac{G_{N}L_{0}}{nr^{n}}+\kappa \theta \frac{r}{n+1}.
\eqnum{3.20}
\end{equation}
But the event horizon is a null surface and satisfies the general
requirement that null-geodesic congruencies have a vanishing acceleration. \
Thus at the event horizon (3.20) takes the form 
\begin{equation}
\kappa \theta _{EH}+\frac{n+1}{n}\frac{G_{N}L_{0}}{\left( r_{EH}\right)
^{n+1}}=0.  \eqnum{3.21}
\end{equation}

Now the Einstein field equations for the $N(N=n+3)$-dimensional space-time
are [4]

\begin{equation}
R_{ab}=8\pi G_{N}\left[ T_{ab}-\frac{1}{n+1}g_{ab}T_{c}^{c}\right] +\frac{%
2\Lambda }{n+1}g_{ab}.  \eqnum{3.22}
\end{equation}
Using equations $\left( 2.3\right) $, $\left( 2.6\right) $, and $\left(
2.8\right) $ with $\left( 3.22\right) $ one finds that 
\begin{equation}
R_{ab}l^{a}l^{b}=\frac{\left( n+1\right) }{n}\frac{G_{N}}{r^{n+1}}\dot{m}%
\left( v\right)  \eqnum{3.23}
\end{equation}

For a spherically symmetric irrotational space-time, such as under
consideration, the vorticity $\omega $\ and the shear $\sigma $ vanish and
the Raychauduri equation (2.10) reduces to 
\begin{equation}
\frac{d\theta }{dv}=\kappa \theta -R_{ab}l^{a}l^{b}-\left( \gamma
_{c}^{c}\right) ^{-1}\theta ^{2}.  \eqnum{3.24}
\end{equation}
Equations (3.21) and (3.23) when substituted in (3.24) show (on neglecting
the term in $\theta ^{2}$) that, indeed, the York condition for the event
horizon is satisfied and we have 
\begin{equation}
\left( \frac{d\theta }{dv}\right) _{EH}\simeq 0.  \eqnum{3.25}
\end{equation}
The event horizons $(EHs)$ in our problem are therefore located by equation
(3.25). Using equation (3.2) equation (3.25) can be written in the form 
\begin{equation}
r^{n+2}-\frac{\left( n+1\right) \left( n+2\right) }{2\Lambda }r^{n}+\frac{%
\left( n+1\right) \left( n+2\right) }{n}\frac{G_{N}m^{\ast }\left( v\right) 
}{\Lambda }=0.  \eqnum{3.26}
\end{equation}
where $m^{\ast }$ is some effective mass given by 
\begin{equation}
m^{\ast }\left( v\right) =m\left( v\right) -\frac{L_{0}}{\kappa } 
\eqnum{3.27}
\end{equation}

Equation (3.26)\ for the location of event horizons is exactly of the same
form as its counterpart equation (3.3) for location of the apparent horizons
with the mass $m$ replaced by the effective mass $m^{\ast }$ as defined in
equation (3.27). Hence borrowing from our previous techniques in solving
equation (3.3) we can immediately write down the solutions to 3.26 as 
\begin{equation}
r_{EH}^{-}\left( v\right) \simeq k\left\{ 1-\frac{1}{4}\left[ 2D^{\ast
}\left( p,\varphi \right) +2R^{\ast }\left( p,\varphi \right) -a^{\ast }%
\right] \right\} ,  \eqnum{3.28}
\end{equation}
and 
\begin{equation}
r_{EH}^{+}\left( v\right) \simeq k\left\{ 1+\frac{1}{4}\left[ 2D^{\ast
}\left( p,\varphi \right) -2R^{\ast }\left( p,\varphi \right) +a^{\ast }%
\right] \right\} ,  \eqnum{3.29}
\end{equation}
where $\ast $\ means $m\left( v\right) \rightarrow m^{\ast }\left( v\right)
=m\left( v\right) -\frac{L_{0}}{\kappa }$ and $\frac{1}{2}\pi <\varphi
^{\ast }<\pi $. In the limit $n\rightarrow 1$ equation (3.28) reduces to, 
\begin{equation}
\mathrel{\mathop{\lim }\limits_{n=1}}%
r_{EH}^{-}\left( v\right) =-\left( \frac{2}{\sqrt{\Lambda }}\right) \cos
\left( \frac{1}{3}\Psi ^{\ast }+\frac{1}{3}\pi \right)  \eqnum{3.30}
\end{equation}
while 
\begin{equation}
\mathrel{\mathop{\lim }\limits_{n=1}}%
r_{EH}^{+}\left( v\right) =\left( \frac{2}{\sqrt{\Lambda }}\right) \cos
\left( \frac{1}{3}\Psi ^{\ast }\right)  \eqnum{3.31}
\end{equation}
where $\left( \frac{\pi }{2}<\Psi ^{\ast }\left( v\right) <\pi \right)
=\arccos \left[ -3m^{\ast }\left( v\right) \sqrt{\Lambda }\right] $. These
limiting cases then reproduce exactly the known equations [21] for locations
of the event horizons in such a four dimensional spacetime. Further, as one
switches $\Lambda $ off one finds from equations (3.17) and (3.3) that the
surface gravity of the black hole measured at the apparent horizon becomes

\begin{equation}
\kappa =\frac{n}{2}\left[ \frac{2G_{N}m\left( v\right) }{n}\right] ^{-\frac{1%
}{n}}  \eqnum{3.32}
\end{equation}
Equation (3.26) along with equation (3.32) imply that in the limit $\Lambda
\rightarrow 0$, then 
\begin{equation}
r_{EH}^{-}\left( v\right) \rightarrow \left( \frac{2G_{N}}{n}\right) ^{\frac{%
1}{n}}\left[ m\left( v\right) -\frac{2}{n}\left( \frac{2G_{N}m\left(
v\right) }{n}\right) ^{\frac{1}{n}}L_{0}\right]  \eqnum{3.33}
\end{equation}
Equation (3.33) locates the event horizon of a radiating black hole imbedded
in a higher dimensional \ Schwarzschild spacetime. And as the dimensionality
is reduced to four, it is clear that 
\begin{equation}
r_{EH}^{-}\left( v\right) \rightarrow 2Gm\left( v\right) \left(
1-4GL_{0}\right) .  \eqnum{3.34}
\end{equation}
This is the exact result originally obtained by York [26] and later verified
by Mallett [18]. It follows then that in equation (3.28) $r_{EH}^{-}\left(
v\right) $\ does indeed represent the locus of the black hole event horizon
in a higher dimensional spacetime with a cosmological constant. Further, the 
$r_{EH}^{+}\left( v\right) $ in equation (3.29) is seen to give the locus of
the cosmological event horizon \ $r_{EH}^{+}\left( v\right) $.

For an observer positioned at $r_{AH}^{-}\left( v\right) <r<r_{AH}^{+}\left(
v\right) $, the region $EH^{-}\cap AH^{-}$ represents the quantum ergosphere
[26]. The ordering of the horizons can now be made. One finds from our
results that $EH^{-}<AH^{-}<AH^{+}<EH^{+}$. The finite location of the
cosmological event horizon $EH^{+}$ is, as can be inferred from equation
(3.29), due to the presence of the cosmological constant. One finds, indeed,
that in the event $\Lambda \rightarrow 0$ then $EH^{+}\rightarrow \infty $.
Again all these results are consistent with the known results for the $N=4$
case.\ 

\section{Structure and Dynamics of the Horizons}

We now turn to the problem of the structure and motion of the various
horizons obtained in our results above.

\subsection{Structure of the Apparent Horizons}

In the foregoing discussion we have found the locations of the apparent
horizons. We now deduce the structure of these surfaces. Since at the
apparent horizons the expansion $\theta $ vanishes then from equations (3.3)
and (2.1) the metric on such surfaces will take the form $%
ds^{2}=2dvdr+d\Omega _{n+1}^{2}$. One finds that equations (2.1), and \
(3.15) as defined by equations (3.13) and (3.14) will induce on the surface $%
r_{AH}^{-}$ a metric of the form 
\begin{equation}
ds^{2}\mid _{r=r_{AH}^{-}}=\frac{k}{2}\alpha _{-}\left( p,\varphi \right) %
\left[ \rho \frac{\sin \frac{1}{3}\varphi }{\sin \varphi }+\sigma \cos \frac{%
1}{3}\varphi \right] \frac{dm}{dv}dv^{2}+d\Omega _{n+1}^{2},  \eqnum{4.1}
\end{equation}
where 
\begin{eqnarray}
\alpha _{-}\left( p,\varphi \right) &=&\left( DR\right) ^{-1}\left[ 2R+\frac{%
\left( 4ac-8c-a^{3}\right) }{4R}-1\right] ,\;\rho =\frac{1}{3}\sqrt{\frac{-p%
}{3}}\frac{d}{dm}\left( \frac{3q}{2\sqrt{\frac{-p^{3}}{3}}}\right) , 
\nonumber \\
\;\sigma &=&\frac{2d}{m\left( v\right) \sqrt{-3p}}.  \eqnum{4.2}
\end{eqnarray}
Noting that $\alpha ,\rho ,\sigma $ are all positive quantities then for $%
\frac{\pi }{2}<\varphi <\pi $, $\frac{dm}{dv}<0$ contributes the only
negative quantity in equation (4.2). We conclude, on this basis, that
according to equation (4.1) the apparent horizon surface $r_{AH}^{-}\left(
v\right) $ of an evaporating black hole in a higher dimensional de Sitter
spacetime is timelike.

Similarly, for $\frac{\pi }{2}<\varphi <\pi $ \ equations (2.1), and \
(3.16) as defined by equations (3.13) and (3.14) will induce on the surface $%
r_{AH}^{+}$ a metric of the form 
\begin{equation}
ds^{2}\mid _{r=r_{AH}^{+}}=-\frac{k}{2}\alpha _{+}\left( p,\varphi \right) %
\left[ \rho \frac{\sin \frac{1}{3}\varphi }{\sin \varphi }+\sigma \cos \frac{%
1}{3}\varphi \right] \frac{dm}{dv}dv^{2}+d\Omega _{n+1}^{2},  \eqnum{4.3}
\end{equation}
where $\rho $ and $\sigma $ are as given in equation (4.2) and 
\begin{equation}
\alpha _{+}\left( p,\varphi \right) =\left( DR\right) ^{-1}\left[ 2R+\frac{%
\left( 4ac-8c-a^{3}\right) }{4R}+1\right]  \eqnum{4.4}
\end{equation}
It follows, then from use of equations \ (4.2) and (4.4) that the
cosmological apparent horizon $r_{AH}^{+}$\ in a higher dimensional
spacetime (as located by equation (4.3)) is spacelike.

In the limit $n\rightarrow 1$, one finds that 
\begin{equation}
ds^{2}\mid _{r=r_{AH}^{-}}=-\frac{4\sin \left( \frac{1}{3}\Psi +\frac{1}{3}%
\pi \right) }{\sin \Psi }\frac{dm}{dv}dv^{2}+d\Omega _{n+1}^{2},  \eqnum{4.5}
\end{equation}
and 
\begin{equation}
ds^{2}\mid _{r=r_{AH}^{+}}=\frac{4\sin \left( \frac{1}{3}\Psi \right) }{\sin
\Psi }\frac{dm}{dv}dv^{2}+d\Omega _{n+1}^{2},  \eqnum{4.6}
\end{equation}
with $\frac{\pi }{2}<\Psi \left( v\right) <\pi =\arccos \left[ -3m\left(
v\right) \sqrt{\Lambda }\right] $.\ \ Equations (4.5) and (4.6) are exactly
the known results [21] for the four dimensional $\left( n=1\right) $ case.

\subsection{The dynamics of the Horizons}

The manner in which the various horizons move can now be inferred. One can
rewrite equations (4.1) and (4.3) in the form 
\begin{equation}
ds^{2}\mid _{r=r_{AH}^{\pm }}\simeq \pm 2L_{0}\Gamma _{\left( AH\right)
n}^{\pm }dv^{2}+d\Omega _{n+1}^{2}  \eqnum{4.7}
\end{equation}
where 
\begin{equation}
\Gamma _{\left( AH\right) n}^{\pm }=\frac{k}{2}\alpha _{\left( \pm \right)
}\left( p,\varphi \right) \left[ \rho \frac{\sin \frac{1}{3}\varphi }{\sin
\varphi }+\sigma \cos \frac{1}{3}\varphi \right] .  \eqnum{4.8}
\end{equation}
We see then that to $O\left( L\right) $\ the apparent horizons $r_{AH}^{\pm
} $\ move with velocities given by 
\begin{equation}
\frac{dr_{AH}^{\pm }}{dv}\simeq \pm 2L_{0}\Gamma _{\left( AH\right) n}^{\pm }
\eqnum{4.9}
\end{equation}

Similarly the motion of the event horizons $r_{EH}^{\pm }$ can be deduced
from equations (3.28) and (3.29). One finds that the velocities of these
surfaces are given by 
\begin{equation}
\frac{dr_{EH}^{\pm }}{dv}\simeq \pm 2L_{0}\Gamma _{\left( EH\right) n}^{\pm }
\eqnum{4.10}
\end{equation}
where the $\Gamma _{\left( EH\right) n}^{\pm }$ are obtained by applying to
equation (4.8) the transformation $m\left( v\right) \rightarrow m^{\ast
}\left( v\right) =m\left( v\right) -\frac{L_{0}}{\kappa }$ that turns
apparent horizon quantities to event horizon quantities. For the range $%
\left( \frac{1}{2}\pi <\varphi ,\varphi ^{\ast }<\pi \right) $ considered
the quantities $\Gamma _{\left( AH\right) n}^{\pm }$ and $\Gamma _{\left(
EH\right) n}^{\pm }$ are positive. It follows then from (4.9) and (4.10)
that for the observer in the region $r_{AH}^{-}<r<r_{AH}^{+}$ both the black
hole horizons $EH^{-}$ and $AH^{-}$ move with respective velocities $%
-2L_{0}\Gamma _{\left( EH\right) n}^{-}$ and $-2L_{0}\Gamma _{\left(
AH\right) n}^{-}$. For such an observer these velocities are negative \
Consequently such motion represents in each case a contraction of the
respective black hole horizon. Conversely, the same equations show that for
the same observer, the cosmological horizons $EH^{+}$ and $AH^{+}$ are
expanding at velocities given by $2L_{0}\Gamma _{\left( EH\right) n}^{+}$
and $2L_{0}\Gamma _{\left( AH\right) n}^{+}$, respectively.

In the event $\left( n\rightarrow 1\right) $ one recovers the known results
[21] for the four dimensional case. And in this limit the results are also
consistent with the usual results that as $\Lambda \rightarrow 0$, we
recover from equations (4.7) and (4.8) the relations 
\begin{equation}
\mathrel{\mathop{%
\mathrel{\mathop{\lim }\limits_{n\rightarrow 1}}}\limits_{\Lambda \rightarrow 0}}%
\frac{dr_{AH}^{-}}{dv}=-2L_{0}  \eqnum{4.11}
\end{equation}
and 
\begin{equation}
\mathrel{\mathop{%
\mathrel{\mathop{\lim }\limits_{n\rightarrow 1}}}\limits_{\Lambda \rightarrow 0}}%
\frac{dr_{EH}^{-}}{dv}=-2L_{0}  \eqnum{4.12}
\end{equation}
\ 

\section{Conclusion}

In this discussion we have examined black holes of the Vaidya type in an
spatially flat higher dimensional spacetime with a cosmological constant.
The analysis revealed the existence of four horizons associated with such a
spacetime and identified as the event horizon, $EH^{-}$ and the apparent
horizon $AH^{-}$ for the black hole, and their cosmological counterparts, $%
EH^{+}$ and $AH^{+}$, respectively. We have pointed out, to good order of
accuracy, the location of these horizons. Further, from our results, we
deduced the structure and discussed the dynamics of these horizons. All our
results reduce to already known results under various limits. In particular,
it was shown at each stage that for the $n=1$, our results reduced exactly
to those previously obtained [21] for the four dimensional case. It is seen
then that the problem of the dynamics of a radiating blackhole in a higher
dimensional cosmological background can be sensibly discussed.

An application of our results to the Hawking radiation problem will be the
topic of a future discussion.

\ 

\ \ 

{\bf Acknowledgments:}

I would like to thank Fred Adams for some insightful discussions and Gordy
Kane, Marty Einhorn and Jean Krisch for their constructive comments.

This work was supported by funds from The University of Michigan.

\ 

\end{document}